\documentclass[conference]{IEEEtran}
\IEEEoverridecommandlockouts
\usepackage{cite}
\usepackage{amsmath,amssymb,amsfonts}
\usepackage{algorithmic}
\usepackage{graphicx}
\usepackage{textcomp}
\usepackage{multicol}
\usepackage{setspace}
\usepackage{xcolor}
\def\BibTeX{{\rm B\kern-.05em{\sc i\kern-.025em b}\kern-.08em
    T\kern-.1667em\lower.7ex\hbox{E}\kern-.125emX}}
\begin{document}

\title{Deep transfer learning for detecting Covid-19, Pneumonia and Tuberculosis using CXR images - A Review\\
}

\author{\IEEEauthorblockN{Irad Mwendo}
\IEEEauthorblockA{\textit{Department of Computer Science } \\
\textit{Dedan Kimathi University of Technology}\\
Nyeri, Kenya \\
iradspm@gmail.com}
\and
\IEEEauthorblockN{Patrick Gikunda}
\IEEEauthorblockA{\textit{Department of Computer Science } \\
\textit{Dedan Kimathi University of Technology}\\
Nyeri, Kenya \\
patrick.gikunda@dkut.ac.ke}
\and
\IEEEauthorblockN{Anthony Maina}
\IEEEauthorblockA{\textit{Department of Computer Science } \\
\textit{Dedan Kimathi University of Technology}\\
Nyeri, Kenya \\
anthony.maina@dkut.ac.ke}
}

\maketitle

\begin{abstract}
\end{abstract}
Chest X-rays remains to be the most common imaging modality used to diagnose lung diseases. However, they necessitate the interpretation of experts (radiologists and pulmonologists), who are few. This review paper investigates the use of deep transfer learning techniques to detect COVID-19, pneumonia, and tuberculosis in chest X-ray (CXR) images. It provides an overview of current state-of-the-art CXR image classification techniques and discusses the challenges and opportunities in applying transfer learning to this domain. The paper provides a thorough examination of recent research studies that used deep transfer learning algorithms for COVID-19, pneumonia, and tuberculosis detection, highlighting the advantages and disadvantages of these approaches. Finally, the review paper discusses future research directions in the field of deep transfer learning for CXR image classification, as well as the potential for these techniques to aid in the diagnosis and treatment of lung diseases.   \\

\section{Background Information}
Due to constant exposure to air particles from the outside environment, human lungs remain highly susceptible to airborne diseases and other injuries \cite{9451726}. These diseases are commonly referred to as Respiratory Diseases (RD) and have become the leading cause of death worldwide \cite{Tan2022}. They include Covid-19 \cite{Liu_2022}, Pneumonia, and Tuberculosis \cite{Bala2021} among others.

Covid-19, is a viral respiratory illness caused by the novel coronavirus SARS-CoV-2, which was discovered in Wuhan, China, in December 2019. The virus spread quickly around the world, resulting in a global pandemic that affected millions of people and caused significant social and economic disruption\cite{Zhou2020}. As of March 8, 2023, there had been over 445 million confirmed cases of COVID-19 worldwide, with over 6 million deaths \cite{WHO_covid19_dashboard}.

Tuberculosis, or TB, is a bacterial infection that primarily affects the lungs but can affect other parts of the body as well\cite{CDC_TB_bacterial_infection}. When an infected person coughs or sneezes, the virus spreads through the air and can cause coughing, fever, and weight loss. TB can be treated with antibiotics, even though the drug resistance is a growing concern. An estimated 10 million people worldwide became ill with tuberculosis in 2020, with 1.5 million dying as a result of the disease\cite{WHO_TB_report_2021} and due to Covid-19, more deaths are expected in 2022 and 2023. 

Pneumonia is another type of respiratory disease that affects the lungs and is caused by a virus, bacteria, or fungi\cite{ALA_pneumonia}. The three types of Pneumonia affects both children and adults even though, viral pneumonia is most common in children below 5 years of age and adults with weak immunity \cite{CDC_viral_pneumonia}. On the other hand, bacterial pneumonia remains the leading cause of mortalities in children and a major cause of illnesses in adults.In United States (US), community acquired pneumonia is estimated to be 4 - 5 million cases annually, with over 1 million hospitalizations and 50,000 deaths\cite{Nicolini_2021}. Fungal pneumonia is less common, and is serious to people with weak immunity. In recent years, fungal pneumonia incidences has increased due to decline in immunity as a result to the growing infections of HIV/AIDS, and Cancer among other factors\cite{ATS_fungal_pneumonia}.

These lung conditions: Covid-19, Pneumonia, and Tuberculosis have serious health consequences and can be fatal if untreated or mismanaged. However, misdiagnosis\cite{wang2023diffuse}, shortage of healthcare professionals\cite{benzaquen2019lung}, and the high cost of treatment\cite{ratnatunga2020rise} continue to be significant challenges. To begin with, lung diseases can be difficult to accurately diagnose, particularly in the early stages, and are frequently misdiagnosed as other respiratory conditions, causing disease progression by delaying proper treatment. Second, there is frequently a shortage of healthcare professionals with specialized training in the diagnosis and treatment of lung diseases, especially in low and middle-income countries. Finally, the cost of treatment for lung diseases, particularly Covid-19, Tuberculosis and Pneumonia, can be high and may not be covered by health insurance, making it difficult for some patients to access necessary care. 

Healthcare technology has become increasingly important and widespread in the medical industry in recent years, particularly in the management and treatment of lung conditions. Chest X-rays, CT scans, and ultrasound are examples of imaging technologies that have been widely used to diagnose and monitor lung diseases. Chest X-rays and CT scans, for example, have been used to diagnose and monitor opacity in COVID-19 patients, while ultrasound has been used to diagnose and monitor pneumonia\cite{kanne2020chest}. Telemedicine has also been widely used during the COVID-19 pandemic for remote consultations, symptom monitoring, and treatment. Patients suffering from lung diseases can now consult with healthcare providers via video conferencing, reducing the risk of transmission and ensuring timely treatment\cite{wurm2008telemedicine}. Finally, Artificial Intelligence (AI) and Machine Learning (ML) have been used to analyze large datasets of patient data, allowing for more precise diagnosis and treatment of lung diseases at a reduced cost\cite{jiang2017artificial}.

\section{Introduction}
Chest X-ray (CXR) images are the most commonly used imaging for the diagnosis of many lung conditions as mentioned in \cite{Liu_2022} and \cite{Bala2021}.CXR images are more preferred to Computed Tomography (CT) scans as they expose the patients to ionizing radiation\cite{Asif2022}. These images requires experts for interpretation\cite{wentzell2018learning}, and as the number of patients suffering from these diseases rises, these experts, who are few, are forced to overwork and in the process, result to delayed diagnoses and timely reporting of results\cite{El-Fiky2021}. Also, due to subjectivity nature of these diseases, experts are likely to produce varying results for same CXR image interpretation\cite{Li2020}.Treatment of lung diseases can be expensive, particularly if one condition is overlooked and requires additional healthcare costs down the line and to overcome such problems, automation for multiple lung diseases detection can help streamline the diagnostic process and enable healthcare providers make more informed decisions about treatment without relying solely on experts. This can help to reduce waiting times, improve patient outcomes, and reduce the burden on radiologists while enabling faster and more efficient diagnoses. This can be achieved with the help of Computer Aided Diagnosis (CAD)\cite{qin2018computer} which can be improved through Computer Vision (CV) and Deep Learning (DL) approaches\cite{gao2018computer}.

In the recent past, DL and traditional machine learning approaches have been widely employed in the classification of medical images into either normal or abnormal instances, allowing experts to specialize in the analysis of abnormal images and classify them accordingly\cite{El-Fiky2021}. These classification are conducted in two steps: the first step involves extracting image features with image descriptors and second step involves feeding those features to a classifier like Support Vector Machines (SVM), and K-Nearest Neighbors (KNN)\cite{Ramteke2012}.The problem with traditional ML approaches is that the accuracy of classification task depends on the manual feature extraction techniques \cite{8575127}.Another approach employed for image classification tasks is DL, which allows automatic feature extraction and classification in one step through use of convolution, pooling and fully connected layers\cite{8575127}. This has been evidenced in the works of \cite{ElAsnaoui2021} to classify Pneumonia images and normal images, and \cite{Liu2019} in detection of lung cancer from normal instances among others.These DL approaches however are computationally expensive\cite{yang2019deep} and suffers from over-fitting and low transfer capability\cite{8575127} and to achieve better results, they necessitates need for large amounts of labelled datasets\cite{doi:10.1177/2472555218818756} which is sometimes, difficult to acquire.

To overcome the challenges with DL, Deep Transfer Learning (DTL), which can also be used in medical image processing is used\cite{misra2020multi}.DTL does not involve training models from scratch, however, it uses pre-trained models similarly to the ones trained on classical DL models like Convolution Neural Network(CNN) to solve one problem, and then, use them to solve different but related problems\cite{krishna2019deep}.These DTL approaches have been widely used in the medical analysis of lung diseases which is evidenced in several works. For instance, in the works of \cite{mamalakis2021denrescov} to automatically classify and detect Covid-19, Pneumonia, and Tuberculosis and similarly, in the works of \cite{katsamenis2020transfer} for the detection of different types of Pneumonia from chest radiographs. Also, ensembled DTL approaches were adopted by \cite{gianchandani2020rapid} to rapidly aid diagnosis of Covid-19 from chest radiographic images which were also used by \cite{boudrioua2020covid}.From most of the works, DTL possess several advantages over DL and traditional ML approaches which are: (1) Requires small dataset for training \cite{huang2017transfer}, (2) High performance \cite{torrey2010transfer} and, (3) less training time\cite{cirecsan2012transfer}. In solving domain problems, DTL adopts strategies like freezing layers\cite{kunze2017transfer}, fine tuning\cite{9241777} and usage of fixed feature extractor (freeze all layers except for final fully connected layer). From the understanding of DTL, the research questions to be addressed are
\begin{enumerate}
    \item What is the role of DTL in the detection of lung diseases?
    \item What are the current state of the art DTL methods used?
    \item What are the advantages of using DTL for lung disease detection?
\end{enumerate}

The rest of the paper is organized as follows: In Section \ref{methodology}, we present the methodological approach followed in filtering relevant state of the art works, in Section \ref{related_works} we present relevant state of the art related works, and in Section \ref{summary} we present the summary and Future works.
\section{Methodology}
The study used a database with information from over 60 scientific articles published between 2016 and 2023 to address the research questions in this domain. These articles were obtained from a variety of sources, including Google Scholar, IEEE Xplore, ScienceDirect, PubMed, and Springer. The articles were chosen based on search phrases related to "Covid-19", "Pneumonia", "Tuberculosis," "Lung diseases", "deep learning", "machine learning", "transfer learning", "deep transfer learning", "Chest x-rays", and other keyword combinations related to lung diseases' detection. The search results were reviewed to select those that were relevant to our work and to eliminate those that were not. The analysis of the articles was done while taking into consideration the lung disease detected in the study, data set used, state of the art method used and the performance obtained based on different evaluation metrics used.
\label{methodology}
\section{Related Works}
\label{related_works}
Many studies have been conducted over the past few years for the detection and automatic classification of different lung conditions. Some of the studies carried out are presented:\\
Panwar et al.\cite{Panwar2020a} proposed a deep transfer learning model nCOVnet made from the original VGG16 model to detect Covid-19 from Chest X-ray images.The only data pre-processing technique used for images was resizing to 224x224x3, and during the training of the model, data leakages were avoided by manually splitting the train and test data set which was evaluated using AUC of ROC, training accuracy and confusion matrix giving an overall accuracy score of 88\%.The data set used was quite small comprising of 337 totals images, with only 192 Covid-19 positive cases.

Three deep transfer learning approaches (ResNet, XCeption and DenseNet) were studied in the detection of patients with Covid-19, Pneumonia, and TB using Chest X-rays\cite{mehta2021classification}. The authors used histogram equalization to minimize bias in the data set, and at the same time, normalized the data. The study achieved high performance accuracy of 98.2\%, 94.21\%, and 93.67\%, obtained in training, validation, and testing sets respectively with ResNet achieving the highest overall performance. However, the data set used was little and imbalanced and comprised of 1229 chest X-ray images with 42 COVID-Medium images, 40 COVID-Mild images, 36 COVID-Severe images, 348 regular images, 263 tuberculosis images, and 500 pneumonia images. 

Oguz et al. \cite{Oguz2022a} used 1345 CT scan images obtained from a research hospital and applied them to different deep learning methods like ResNet-50, ResNet-101, AlexNet, GoogleNet, and other classification methods like SVM, Random Forest, Decision Trees, etc. to reduce diagnosis time for COVID-19. Pre-processing activities involved reducing the dimensions of the CT images through maximum pooling and applying the ReLU activation function to each convolution layer. The experimental results have shown that ResNet-50 and SVM performed better than other models with 96.3\% accuracy and an F1-score of 95.87\%. This work is however limited due to the size of data used and might not generalize well to new data.

Kassania et al.\cite{Kassania2021a} compared different deep transfer learning approaches (MobileNet, DenseNet, Xception, ResNet, InceptionV3, InceptionResNetV2, VGGNet, NASNet) in the automatic prediction of COVID-19 Pneumonia cases in Chest x-rays and CT scans. The images used were resized and normalized. The data set had 274 images with 137 images of positive Covid-19 cases and 137 images of health patients.Densenet-121 achieved high performance with a classification accuracy of 99\% followed by ResNet-50 with 98\% classification accuracy. Despite the high performance, the study focused on detecting two conditions i.e. Covid-19 and Normal instances, and also, used little data set.

The more classes used the lower the accuracy of the model. This was evidenced in the works of Hussain et al. \cite{Hussain2021a} who proposed a deep CNN model 'CoroDet' with 22 layers. The total data set used had a total of 7390 images where 2843 were for COVID-19 case,3108 for normal cases and 1439 for pneumonia (both viral and bacterial) cases. Without pre-procesing the data, they applied their deep CNN model to 2 (Covid-19 ,Pneumonia), 3(Covid-19, Pneumonia-Bacteria, Normal), and 4(Covid-19, Pneumonia-Bacteria, Normal, Pneumonia-Viral) class classifications.Their model achieved an accuracy of 99.1\%, 94.2\% and 91.2\% for 2, 3 and 4 class classifications respectively. However, these results can be improved by increasing the data, balancing the class instances and pre-processing the data to minimize noise. 

In the works of Haritha et al.\cite{Haritha2020a}, pre-trained models appeared to perform better in the detection of Covid-19  cases. The researchers developed CheXNet model from the original DenseNet121 to detect any anomalies in the chest x-ray dataset which was re-scaled, and augmented through flipping, rotating, and zooming to increase the size of the training set leading to achieving an accuracy score of 99.9\%. This was however limited due to use of small data sets which comprised of 1824 Chest x-ray images (912 for Covid-19 cases and 912 for Non-Covid cases).

Sai et al.\cite{SAIKOUSHIK2021109153} proposed a deep transfer learning CNN method for detecting viral Pneumonia, bacterial Pneumonia, and Covid-19. Using a pre-trained ResNet-50V2 model, chest radiographs(5856 images) were trained on 100 epochs resulting in accuracy and test score of 94\%. The high accuracy was achieved through an increase in the number of epochs and the addition of a dropout layer after the dense layer.

In an attempt to classify whether a person has lung disease or not, Bharati et al.\cite{Bharati2020a} developed a hybrid deep-learning model using CNN, VGG16, data augmentation, and spatial transformer networks. Using attributes of age, chest x-ray images, a person’s gender, and view position, their hybrid model obtained a validation accuracy of 73\% despite the noisy and complex data used. To improve detection accuracy, Rajagopal et al.\cite{Rajagopal2023a} proposed the usage of an Arithmetic Optimization algorithm (AOA) on deep CNN and further reduces the noise of images by applying an “Anisotropic Diffusion Filter Based Unsharp Masking and crispening scheme”. Lung Cancer, Tuberculosis, and Pneumonia are also detectable through imaging, Khobragade et al.\cite{Khobragade2016a} proposed an automatic detection of these diseases using an artificial feed-forward neural network. To ensure better performance of the model, the histogram equalization technique was used in the improvement of image intensity and contrast. This resulted in an accuracy of 92\%. An ensemble learning method of random forest, SVM, and Logistic regression was utilized by Ravi et al.\cite{Ravi2022a} in the lung disease detection and achieved an accuracy of 98\% while Bhosale et al.\cite{Bhosale2023a} developed an ensemble deep transfer learning CNN model for the detection of obstructive pulmonary diseases with an accuracy of 99.70\%.

Artificial Neural Networks (ANN) aids in the detection of respiratory diseases diagnosed with chest radiographs. Usha et al.\cite{UshaKiruthika2019a} developed a model for Tuberculosis detection. The contraction path technique was used with 4 blocks each having convolution layers, activation functions, and max pooling. This resulted in an accuracy of 81\%. Non-Linear and Fisher's discriminant techniques were adopted by Seixas et al.\cite{Seixas2013a} to detect pleural Tuberculosis based on clinical grounds.  Two Deep transfer learning CNN networks, AlexNet and GoogleNet were used in the classification of chest radiographs to detect pulmonary Tuberculosis according to Lakhani et al.\cite{Lakhani2017a} with radiologists assisting in cases where discrepancies occurred. Ahsan et al.\cite{Ahsan2019a} applied deep transfer learning model of VGG 16 to detect whether patients have Tuberculosis or not. Sigmoid activation was used in the output layer while Adam optimizer was used to update weighs resulting in an accuracy of 80\% in the model. Liu et al.\cite{Liu2017a} used shuffle sampling technique to deal with the unbalanced data set in the classification of different TB manifestations, and achieved a classification accuracy of 85.68\%.

\section{Summary and Future works}
\label{summary}
The reviewed research works demonstrate the efficacy of employing deep transfer learning approaches in the detection of lung diseases. Deep transfer learning models mostly used include ResNet-50, VGG-16, InceptionV3, ResNet-18 and DenseNet-121 among others \cite{SAIKOUSHIK2021109153}\cite{Bharati2020a}\cite{minaee2020deep}. 

To improve the performance of models, the researchers have utilized techniques like transfer learning, data augmentation and hyper-parameter tuning\cite{minaee2020deep}\cite{apostolopoulos2020covid}. TL has been used as the base step to allow training of the pre-trained models on new datasets \cite{apostolopoulos2020covid}. Data augmentation techniques have been applied to generate new training samples by applying transformations to the existing data set \cite{minaee2020deep}\cite{chowdhury2020can}.Hyperparameter tuning optimized the parameters that control the behavior of deep learning models, such as the learning rate or the number of layers \cite{minaee2020deep}\cite{zhang2021automated}.
Despite the advancements of research in this field, there are still pre-existing gaps:
\begin{enumerate}
    \item The studies reviewed relatively use small data sets, for instance\cite{Panwar2020a}.
    \item Most of the studies reviewed focuse on detecting one lung disease
    \cite{Panwar2020a}\cite{Oguz2022a}.
     \item In the experimental analysis of different deep and transfer learning approaches, data imbalance\cite{Liu2017a} issue remains a key problem and this could result to biased results obtained in many researches.  
\end{enumerate}

\vspace{5pt}
\begin{onehalfspacing}
		\bibliographystyle{IEEEtran}
		\bibliography{References}
\end{onehalfspacing}
\vspace{12pt}

\end{document}